\documentclass[12pt,preprint]{aastex}

\newcommand{\beq}{\begin{equation}}
\newcommand{\beqa}{\begin{eqnarray}}
\newcommand{\eeq}{\end{equation}}
\newcommand{\eeqa}{\end{eqnarray}}

\begin{document}

\title{A feedback compression star formation model and the black hole-bulge relations}

\author{Bing-Xiao Xu$^1$ and Xue-Bing Wu$^1$}

\affil{$^1$Department of Astronomy, Peking University, 100871
Beijing, China} \email{xubx@bac.pku.edu.cn, wuxb@bac.pku.edu.cn, }

\begin{abstract}
We present a "feedback compression" model to describe the galactic spheroid
formation and its relation with the central nuclear activity. We
suggest that the star formation itself can serve as the "positive
feedback" in some extremely dense region to trigger the starburst.
The star formation rate as well as the related stellar
feedback-induced turbulence will be maximized under the regulation
of the background dark halo's gravity. There is also stellar
feedback acting inward to confine and obscure the central black hole (BH) till
the BH grows sufficiently large to satisfy a balance
condition between the accretion disk wind and the inward stellar feedback.
The extremely vigorous star formation activity, the
BH-bulge relations, the maximum velocity dispersion as well as the
maximum BH mass are investigated based on such scenario, and are found to be
consistent with observations.
\end{abstract}


\section{INTRODUCTION}
Observations have shown clear evidence that the mass of
supermassive black hole (SMBH) in the center of every galaxy is
tightly correlated with the velocity dispersion of the bulge stars
\citep{FM00,GB00,TMA02} and the mass of the whole bulge
\citep{MA98,MD02,MH03}, which reflect the interplay with the
nuclear activity and the spheroidal star formation activity
(potential well depth of bulge stars). Recent consensus emphasizes
that the nuclear black hole (BH) feedback is the sticking point to
explain these correlations: the central BH can interact with the
surrounding environment through BH feedback in a self-regulated
way. However, the tightness of the correlations indicated by both
the observations and numerical simulations pose big challenge to
many conventional feedback models, whose results are sensitive to
the variable parameters such as gas fraction (the mass fraction of
gas to dark matter) \citep{DM05}. Recently, some modified feedback
models have been proposed to alleviate the inconsistency.
\citet{BN05} suggest that the BH accretion physics on local scale
(relative to previous "global" models) should be considered. They
consider that the "maximized" accreted gas mass will make the
result insensitive to the gas fraction. \citet[][hereafter
XWZ07]{XWZ07} provide another scenario to explain both the
starburst activity in protospheroid and the BH-bulge relations.
They suggest that the spheroid formation is "halo-regulated":  the
star formation is regulated by the background dark matter halo. In
particular, the strong star-forming feedback acts outward to
resist the gravity from the dark halo while acts inward to feed
and obscure the central BH. Such scenario can naturally link the
nuclear activity and outside star formation activity, and well
explain the BH-bulge relations.

Here we revisit the idea of XWZ07 and find the "halo-regulated"
mechanism can be generalized and simplified. We argue that it is
possible that the stellar feedback can compress the surrounding
gas and trigger the catastrophic star formation activity in a very
dense environment when neglecting the heating effects. The intense
starburst will make the star-forming region highly turbulent. The
"maximized" turbulent sound speed, which is regarded as a measure
of potential well depth of the protospheroid, can be reached when
the whole system is in the virial equilibrium. Based on such "feedback
compression" model, we show that once the assumption of homogeneous
turbulent environment is valid, the star formation is regulated by
the dark halo's gravity and regulates the central BH growth. The
resultant BH-bulge relations are universal: only related to the
velocity dispersion of stars and dark matter profile.

\section{STAR FORMATION UNDER THE MOMENTUM FEEDBACK COMPRESSION}
The merger induced starburst regions have very dense environment:
the molecular gas density can be as high as $10^3\sim10^4\rm
cm^{-3}$ \citep{DS98}, much denser than that of disk galaxies. The
cooling by collisionally excited atomic and molecular emission
processes can be very efficient. Hence, we hypothesize that the
heating feedback due to the ionization of HII region may be
neglected comparing with the mechanical feedback in the starburst
region. Since the shock-heated gas can quickly radiate their
thermal energy, it is possible that the outward propagating shock
can compress the surrounding gas into a gas shell in a momentum
driven way and trigger the new star formation events. The result
can be understood by a simple example: assuming that the expanding
gas shell is thin, its dynamics can be described by \beq{d \over
dt}\left(R^3{dR \over dt}\right)={3\dot{M}_wv_w \over
4\pi\rho_0},\eeq where $\dot{M}_w$ is the mass loss rate, $v_w$ is
the velocity of wind and $\rho_0$ is the density of homogeneous
gas around the stars. We can easily find the following solution at
large $R$ \beq R(t)=\left({3\over2}{\dot{M}_wv_w \over
\pi\rho_0}\right)^{1/4}t^{1/2}.\eeq The shell will finally stall
due to the ambient pressure, and we have the stalling radius
$R_s=(\dot{M}_wv_w/ 4\pi n_0kT)^{1/2}$. Using Eq. (2), the total
compression timescale can be estimated as \beq t_c={\mu
m_{H_2}\over 4kT}\left({2\dot{M}_wv_w \over
3\pi\rho_0}\right)^{1/2}.\eeq Comparing $t_c$ with the dynamical
timescale $t_{dyn}=(3\pi/32G\rho)^{1/2}$, we have \beq {t_c\over
t_{dyn}}\approx5\left({\rho_c\over\rho_0}\right)^{1/2}\dot{M}_6^{1/2}v_{500}^{1/2}T_{10}^{-1},\eeq
where $\rho_c$ is the gas density of the compressed gas shell,
$\dot{M}_6=\dot{M}_w/10^{-6}M_{\odot} yr^{-1}$, $v_{500}=v_w/500km
s^{-1}$ and $T_{10}=T/10K$. Note that the shell gas density
$\rho_c$ is larger than $\rho_0$, so $t_c$ is at least one order
larger than $t_{dyn}$. It demonstrates that in a very dense
environment, the density inhomogeneities are amplified by the
feedback compression and the gas has enough time to collapse into
the cloud.

The "light" outflow accelerating against the dense cold gas shell
will trigger the Rayleigh-Taylor instability (RTI). Moreover, the
Kelvin-Helmholtz instability (KHI) may also occur as the outflow
punches into the gas shell and moves through the cold gas. The
combination of these two instabilities will prevent the gas collapse
and destruct the star-forming cloud. If the density contrast is
large, the dispersion relation for RTI is $w\approx(ka)^{1/2}$
where $a$ is the acceleration \citep{CK61}. The characteristic
growth timescale for RTI responsible for the cloud destruction is
$\tau_{RT}\approx(2\pi l_J/a)^{1/2}$ where $l_J$ is the Jeans length.
The acceleration can be derived from Eq. (2) \beq
a(t)={1\over4}\left({3\over2}{\dot{M}_wv_w \over
\pi\rho_0}\right)^{1/4}t^{-3/2}\eeq Initially, the growth
timescale $\tau_{RT}$ may be smaller than the dynamical timescale
$t_{dyn}$. The RTI will make the dense cold shell highly porous to
the hot wind and entrain cold gas into the wind. The efficiency of
driving wind enhances \citep{SL01}. Note $\tau_{RT}\propto
t^{3/4}$, we define the duration in which RTI dominates by setting
$\tau_{RT}=t_{dyn}$. Using Eq. (5), the duration is \beq
t_d=2.5\times10^5
\dot{M}_6^{1/6}v_{500}^{1/6}n_3^{-1/2}T_{10}^{-1/3}\left({\rho_c
\over\rho_0}\right)^{-1/3}yr, \eeq where $n_3=n_0/10^3 cm^{-3}$.
The duration is smaller than the dynamical timescale $t_{dyn}$,
which indicates that the cloud destruction by RTI and KHI (its
growth timescale is of the same order as RTI's \citep{AG06}) may not
be important in the momentum driven phase, at least at large radius
$R$. So it is reasonable to assume that the feedback induced
compression can trigger and expedite the cloud collapse and the
star formation.

\section{"HALO-REGULATED" STAR FORMATION AND BLACK HOLE GROWTH}

Star formation in the local disk galaxies is usually
inefficient and the typical star formation efficiency (SFE) is
$\sim2\%$ \citep{KN98}. It is because that the star formation is
regulated by the "negative" feedback (heating, blown out) to keep
the star formation rate (SFR) from raising too high. However, as
mentioned in the above section, it may not be the case in
the starburst region. Without the heating effects, the feedback
induced compression can serve as a "positive" feedback to
trigger the intense star formation activities unless the total
mechanical energy generated by the stellar feedback is larger than
the binding energy of these regions. In another words, once the
total feedback from those massive stars are unable to disrupt or
unbind the whole star forming region, the star formation process
may continue for a relatively long time, which is analogous to the
formation of bound clusters \citep{EE97}.

Following XWZ07, we adopt the NFW density profile for the background dark
matter and assume the standard cosmological parameters
with $\Omega_m=0.3,\Omega_{\Lambda}=0.7$ and $h=0.7$ ($z=0$). The
inner NFW profile is given by \beq \rho_{NFW}(r)\approx \Pi
r^{-1}, \qquad \Pi \equiv 130 M_{\odot} {\rm pc}^{-2}
M_{v,12}^{0.07}[\xi(z)]^{2/3}\Psi_{c,0.58}^{-1}\eeq
\citep{NFW97,KS01}, where $M_{v,12}=M_{vir}/10^{12}M_{\odot}$ and
$M_{vir}$ is the virial mass of the halo,
$\xi(z)=[(\Omega_m/\Omega_m(z))(\Delta_c/100)],\Omega_m(z)=[1+(\Omega_{\Lambda}/\Omega_m)(1+z)^{-3}]^{-1},\Delta_c=18\pi^2
+82d-39d^2$ where $d=\Omega_m(z)-1$ \citep{BN98,BL01}.
$\Psi_{c,0.58}=[\ln(1+c)-c/(1+c)]^{-1}/0.58$ where
$c\approx13.4M_{v,12}^{-0.13}(1+z)^{-1}$ is the concentration
parameter \citep{BU01}. Such $r^{-1}$ profile in the inner region
is almost a universal profile: it doesn't change by the galaxy
merger or interaction, it is also insensitive to the redshift and
the halo mass. Another interesting thing to note is that such
inner profile gives a nearly constant gravitational acceleration
\beq g_{\rm DM}(r)=\frac{GM_{DM}(r)}{r^2} = 2 \pi G \Pi \sim 1.2
\times 10^{-8}{\rm cm}\ {\rm sec}^{-2}. \qquad \eeq

Intense star formation activity and related stellar feedback will
make the protospheroid environment highly turbulent.
Because the bulge stars density distribution implies that the protospheroid
density profile may follow $\rho\sim r^2$ \citep{TMA94}, here we
assume an isothermal density profile for the protospheroid baryon
or mixture of gas and stars \beq\rho_b(r)={c_s^2\over2\pi
Gr^2},\qquad M_b={2c_s^2r\over
 G},\eeq where $\rho_b(r)$ is the total baryon density or density of the
mixture of gas and stars, $c_s$ is the "isothermal" turbulent
sound speed and $M_b$ is the enclosed mass within the radius $r$.
The momentum transport during the compression requires that \beq
{\dot{P}_{\star}\over4\pi r^2}=\rho_b(r)c_s^2,\eeq where
$\dot{P}_{\star}$ is the net outward momentum deposition rate. The
star formation activity and the feedback-induced turbulence will
be maximized
 till the whole system evolves to a virial equilibrium state. Once the
 whole system becomes a little more turbulent, the deposited momentum flux
 will make the whole system deviate from the
equilibrium state and the SFR is hence avoided from raising
higher. Such a self-regulated mechanism will make the whole system
maintain the equilibrium state.

 We can write the equation of the virial
 equilibrium for the protospheroid as
 \beq3M_bc_{s,m}^2={GM_b^2\over r}+\pi\Pi GM_br,\eeq where $c_{s,m}$
 is the maximum turbulent sound speed.
  The first term of the right side of Eq. (11) denotes the total baryon's
  self-gravity while the second term denotes the
 gravity from the background dark matter.
 Substituting Eq. (9) into Eqs. (10) and (11), we obtain
 \beq\dot{P}_{\star}={2c_{s,m}^4\over G}=\pi\Pi GM_b.\eeq
 Eq. (12) shows that the maximum turbulent sound speed or the
 potential well depth of the protospheroid is directly related to
 the background dark matter. In another words, both the star formation
 activity and the feedback compression are regulated by the dark
 halo's gravity as XWZ07 proposed. It only requires the homogeneous turbulent environment
 and the virial equilibrium, without involving the detailed gas
 assembly physics such as monolithic collapse or merger driven
 inflow. So the feedback compression combined with the halo regulation
 scenario provides a more general description to the star formation process
 during the spheroid formation.

 Following XWZ07, during the formation of the protospheroid, the stellar feedback can act in both inward and
 outward directions. At small scale (e.g. galactic nuclear region), the
 inward stellar feedback (required to conserve the local momentum)
 obscures and regulates the BH growth, while the outward stellar feedback
 resists the gravity at large scale. In particular, the inward stellar feedback
 regulates the BH growth by interacting with the Compton-thick wind
 launched from the accretion disk if super-Eddington accretion is
 assumed \citep{KP03}. The final balance between the inward stellar feedback and
 the disk wind is achieved when
\beq \dot{P_{\star}}=\frac{8\pi GM_{BH}}{\kappa} \eeq where
$\dot{P_{\star}}$
 is the momentum flux transported by inward stellar feedback and the
 right side of Eq. (13) is the momentum deposition of the disk wind.
 Then if the BH's
feedback is large enough to halt the further gas supply, and a BH will end
its main growth phase after Eq. (13) is
satisfied.

 At a late epoch of the coeval evolution, star formation consumes most of gas and gradually fades away.
 Without continuous ejecting energy and momentum from the stellar feedback, the feedback-induced turbulence will decay on a crossing time of
the system \citep{ST98}.
 Then the remaining stellar system will be virialized through violent relaxation under the combined
 gravity from itself and the background dark matter. We call the system  at  ''initial state'' after turbulence
 decay and before virialization, and at ''final state'' after virialization. Assuming the total energy
 of the initial state is $E$, the kinetic energy of the final virialized state is $K=-E$
 according to the virial theorem. So we have
 \beq3\sigma_f^2=2\left({GM_b\over r}+\pi\Pi Gr\right)=6c_{s,m}^2.\eeq
 Using Eq. (12), we obtain
 \beq\dot{P}_{\star}={\sigma_f^4\over 2G},\qquad r_b={\sigma_f^2\over2\pi\Pi
 G},\eeq where $r_b$ is the boundary radius of the initial state.
 We note that the total momentum deposition rate from stars is
 only related to the velocity dispersion of the final stellar
 system, independent of any parameters of the detailed star formation
 physics.

Using Eqs. (13) and (15), we obtain the final BH mass \beq
M_{BH}^{final}=\frac{\kappa\sigma_f^4}{16\pi
G^2}=1.5\times10^8\sigma_{200}^4M_{\odot}.\eeq The result is
remarkably consistent with the low-redshift observations
\citep{TMA02}. The stellar bulge mass is approximately equal to
$M_b$. Using Eqs. (12) and (13), the ratio of BH mass to bulge
 mass can be expressed as
\beq\frac{M_{BH}}{M_{bulge}}=\frac{\kappa\mathit{\Pi}}{8}\approx1.4\times10^{-3}M_{v,12}^{0.07}[\xi(z)]^{2/3}\Psi_{c,0.58}^{-1},\eeq
 which matches the Magorrian relation found for the local galaxies
\citep{MA98,MD02,MH03}.

\section{APPLICATION TO THE HIGH-REDSHIFT STAR FORMING GALAXIES}

The gas fraction of high-redshift galaxies is much larger than
that of local galaxies. In an extremely case, we take the fraction
$\sim1$. In another words, the protospheroid with mass of $M_b$ is
almost totally in the gaseous form. Large amount of gas
accumulating in the central region will trigger the central
vigorous starburst accompanying with strong star-forming feedback.
We mainly focus on two primary sources of star forming feedback:
radiation pressure and supernovae. The combined momentum flux
deposited in these star-forming feedback can be written as
\citep{MQT05,XWZ07}\beq \dot P_{\star}=\dot P_{rp}+\dot
P_{sn}=\xi_m\epsilon\dot M_{\star}c,\eeq where $\dot M_{\star}$ is
the star formation rate and $\xi_m=1+\dot P_{sn}/\dot P_{rp}$ is
of the order of unity in our model. Combining Eqs. (13), (16) and
(18), we can easily obtain  the star formation rate of  high
redshift starburst galaxies \beq \dot
M_{\star}=\frac{\sigma^4}{2G\xi_m\epsilon
c}\approx600\sigma_{200}^4\xi_m^{-1}\epsilon_3^{-1}M_{\odot}yr^{-1}.\eeq
Although the star formation law in the protospheroid is far from
clear, we adopt an equivalent Schmidt-Kennicutt Law in order to
compare with the local disk galaxies \citep{SM59,KN98}\beq\dot
M_{\star}=\eta\frac{M_b}{t_{dyn}},\eeq where $\eta$ is the
equivalent star formation efficiency and
$t_{dyn}=(3\pi/32G\rho_g)^{1/2}$ is the dynamical time scale. In
our model we take $\rho_g=3M_b/4\pi r_b^3$ as the average gas
density where $r_b$ is the outer boundary radius given in Eq.
(15).

From Eqs. (12) and (18), we have \beq\xi_m\epsilon\dot
M_{\star}c=\pi\Pi GM_b.\eeq Using Eq. (20) to eliminate $\dot
M_{\star}/M_b$ in Eq. (21), we get the equivalent star formation
efficiency as \beq \eta=\frac{\sqrt{2}\sigma_f\pi}{8\xi_m\epsilon
c}=0.4\sigma_{200}\xi_m^{-1}\epsilon_3^{-1},\eeq where
$\epsilon_3=\epsilon/10^{-3}$.

 We find that our derived equivalent SFE is
much higher than that in normal disk galaxies inferred from the
Kennicutt Law, but is consistent with the high redshift star
formation observations and some small scale star formation
observation (eg. SFE in the formation of the protocluster). We note
that the derived SFE is independent on the radius of the
star-forming region and time, and the larger the velocity
dispersion is, the higher the star formation efficiency is. Eq.
(22) also gives us another implication for the maximum stellar
velocity dispersion. We can rewrite Eq. (22) as
\beq\sigma_f=\frac{4\sqrt{2}\xi_m\eta\epsilon c}{\pi}.\eeq  The
physical limit requires $\eta\le1$, so the maximum stellar
velocity dispersion is
\beq\sigma_f^{max}=\frac{4\sqrt{2}\xi_m\epsilon
c}{\pi}=540\xi_m\epsilon_3 km s^{-1}.\eeq According to Eq. (16),
the maximum BH mass is $8\times10^9M_{\odot}$.

Using Eqs. (12), (15) and the expression of $t_{dyn}$ , we can also
obtain the characteristic star formation timescale \beqa
t_{\star}&=&\frac{t_{dyn}}{\eta}=\frac{2r_b\xi_m\epsilon
c}{\sigma_f^2}={\xi_m\epsilon c\over\pi\Pi G}\nonumber\\
&\approx&1.0\times10^8\xi_m\epsilon_3M_{v,12}^{-0.07}[\xi(z)]^{-2/3}\Psi_{c,0.58}yr.\eeqa

Through the feedback compression, the dark halo's gravity regulates
the SFR and SFE to a higher level during the spheroid formation.
We note that for some luminous elliptical galaxies whose velocity
dispersion $\sigma\sim300kms^{-1}$, the predicted star formation
rate can reach as high as $3000M_{\odot}yr^{-1}$, which is
consistent with the observations of high-redshift starburst
galaxies \citep{SV05}. It is also interesting to note that the
star formation time scale is independent with velocity dispersion
$\sigma$. Furthermore, the characteristic star formation time
scale is larger than the salpeter time scale, which is usually taken to be
the typical time scale
for BH's growth. This means that the BH
will first grow relatively fast to reach the balance condition (Eq. (13)) and
then grow relatively slow to response the outside stellar
feedback.

\section{DISCUSSIONS}
Unlike the previous momentum feedback models
\citep{KI03,MQT05,BN05}, which mainly focus on the BH feedback
dynamics, our scenario considers more about the star formation
activity in the protospheroid and its relation with the nuclear BH
growth. Our model favors gas-rich environment at high-redshift
because large amount of gas is needed to form stars and obscure
the BH. We argue that at early epoch of BH growth (when BH is relatively
small),
large scale outflow or jet may not be crucial in producing
BH-bulge relations although they become important after BH's main
growth epoch \citep{CH02}. In another words, BH doesn't generate
outflow or jet till the balance condition Eq. (13) is satisfied.
The derived $M_{BH}-\sigma$ relation based on our model is insensitive
to the gas
fraction and other variables, because the velocity
dispersion is the measure of the maximized turbulent velocity of
the total baryon component rather than the gas only, as Eq. (9)
shows. The derived $M_{BH}-M_{bulge}$ relation has weak dependences
on the redshift and the halo mass, which offer the intrinsic
scatters to the relation.

Extremely high SFR and SFE are the results of certain "positive"
feedback which is probably due to either "internal" or "external"
effects. \citet{SL05} suggest that the high SFR and SFE are
triggered by the super-Eddington outflow driven by the SMBH. Such
"external" positive feedback naturally leads to a top-heavy
initial mass function (IMF) which is preferred by the early
generation of star formation and predicts an antihierachical trend
of SMBH growth \citep{MRD04}. However, recent optical, infrared
and X-ray studies of SMGs indicate that SMGs harbour relatively
smaller SMBH than that of typical quasars
\citep{IV98,VC01,SL03,AX05} and the main growth phase of SMBH
("pre-quasar" phase) is heavily obscured. Considering the SMGs
themselves are massive galaxies which are reckoned as the
progenitors of local ellipticals \citep{GR05}, the vigorous star
formation activity can not be regulated by the small BH.
Reversely, the star formation activity should have great impact on
the small BH. So we argue that the "internal" positive feedback
 which is produced by the star formation itself may
exist in some extremely dense starburst region at median redshift,
and the SMBH outflow triggered star formation mode may only be
available at very high redshift \citep{WT04}. In addition, the
maximized "positive" stellar feedback is actually related to the
dark halo's gravity in our model. Under the regulation of dark
halo's gravity, the maximized velocity dispersion also has its
maximum value, which is determined by the physical upper-limit of
SFE. The physical upper-limit of BH mass ($\sim10^{10}M_{\odot}$)
can then be obtained by the $M_{BH}-\sigma$ relation, and the
dynamical signature of such SMBHs should be detectable
\citep{WL03}. We note that some observational evidence do support
such result \citep{NT03,VG04}, although all of them still contain
a lot of uncertainties. We expect more accurate SMBH mass
measurements in the future to confirm our result.

\section{Acknowledgments}
XBW acknowledges the support from NSFC grants (No.10473001 and
No.10525313), RFDP grant (No.20050001026) and Program for New
Century Excellent Talents in Universities of China.

\end{document}